\documentstyle[preprint,epsf,eqsecnum,aps,floats,tighten]{revtex}

\def\Missing#1#2{{\mbox{$#1\kern-0.57em\raise0.19ex\hbox{/}_{#2}$}}}

\def\vMissing#1#2{\ifmmode
            \vec{#1}\kern-0.57em\raise.19ex\hbox{/}_{#2}
         \else
            {{\mbox{$\vec{#1}\kern-0.57em\raise.19ex\hbox{/}_{#2}$}}}
         \fi}
\def\lsim{\mathrel{\rlap{\lower4pt\hbox{\hskip1pt$\sim$}}
    \raise1pt\hbox{$<$}}}        
\def\gsim{\mathrel{\rlap{\lower4pt\hbox{\hskip1pt$\sim$}}
    \raise1pt\hbox{$>$}}}        

\def\met{\mbox{$\Missing{E}{T}$}}

\def\D0{D\O }

\def\simge{\mathrel{\rlap{\raise 0.53ex \hbox{$>$}}%
{\lower 0.53ex \hbox{$\sim$}}}}
\def\simle{\mathrel{\rlap{\raise 0.53ex \hbox{$<$}}%
{\lower 0.53ex \hbox{$\sim$}}}}

\def\ETmiss{\mbox{${\hbox{$E$\kern-0.5em\lower-.1ex\hbox{/}\kern+0.15em}}_T$ }}

\def\err#1#2#3 {{\it Erratum} {\bf#1},{\ #2} (19#3)}
\def\ib#1#2#3 {{\it ibid.} {\bf#1},{\ #2} (19#3)}
\def\nc#1#2#3 {Nuovo Cim. {\bf#1} ,#2(19#3)}
\def\nim#1#2#3 {Nucl. Instr. Meth. {\bf#1},{\ #2} (19#3)}
\def\np#1#2#3 {Nucl. Phys. {\bf#1},{\ #2} (19#3)}
\def\pl#1#2#3 {Phys. Lett. {\bf#1},{\ #2} (19#3)}
\def\prev#1#2#3 {Phys. Rev. {\bf#1},{\ #2} (19#3)}
\def\prl#1#2#3 {Phys. Rev. Lett. {\bf#1},{\ #2} (19#3)}
\def\rmp#1#2#3 {Rev. Mod. Phys. {\bf#1},{\ #2} (19#3)}
\def\zp#1#2#3 {Zeit. Phys. {\bf#1},{\ #2} (19#3)}     

\begin{document}

%
%
\title{Search for Leptoquark Pairs Decaying to $\nu\nu ~+~jets$ in $p{\bar p}$
Collisions at $\sqrt{s}$=1.8 TeV
}
\author{\centerline{The D\O\ Collaboration
  \thanks{Submitted to the {\it International Europhysics Conference
        on High Energy Physics},
	\hfill\break
	July 12-18, 2001, Budapest, Hungary,
        \hfill\break 
	and  {\it XX International Symposium on Lepton and Photon Interactions at High Energies}
	\hfill\break
        July 23 -- 28, 2001, Rome, Italy. 
	 }}}
\address{
\centerline{Fermi National Accelerator Laboratory, Batavia, Illinois 60510}
}
%
%
\date{\today}

\maketitle

%
%
\begin{abstract}
We present the preliminary results of a search for leptoquark ($LQ$) pairs  
using (85.2  $\pm$ 3.7) pb$^{-1}$ of $p{\bar p}$ collider data collected by  
the D\O\ experiment at the Fermilab Tevatron from 1994-1996.  We observe no  
evidence for leptoquark production and set a limit on $\sigma (p{\bar p}  
\rightarrow LQ\overline{LQ} \rightarrow \nu\nu~+~jets)$ as a function of the  
mass of the leptoquark ($M_{LQ}$), assuming BR($LQ \rightarrow \nu  
q$)=100\%.  At the 95\% confidence level, we exclude scalar leptoquarks for  
$M_{LQ} <$ 99 GeV/c$^2$, and vector leptoquarks for $M_{LQ} <$ 178 GeV/c$^2$. 

\end{abstract}

\newpage
\begin{center}
%
V.M.~Abazov,$^{23}$                                                           
B.~Abbott,$^{58}$                                                             
A.~Abdesselam,$^{11}$                                                         
M.~Abolins,$^{51}$                                                            
V.~Abramov,$^{26}$                                                            
B.S.~Acharya,$^{17}$                                                          
D.L.~Adams,$^{60}$                                                            
M.~Adams,$^{38}$                                                              
S.N.~Ahmed,$^{21}$                                                            
G.D.~Alexeev,$^{23}$
% ...
\\                                                                            
\vskip 0.30cm                                                                 
\centerline{(D\O\ Collaboration)}                                             
\vskip 0.30cm                                                                 
\centerline{$^{1}$Universidad de Buenos Aires, Buenos Aires, Argentina}       
\centerline{$^{2}$LAFEX, Centro Brasileiro de Pesquisas F{\'\i}sicas,         
                  Rio de Janeiro, Brazil}                                     
\centerline{$^{3}$Universidade do Estado do Rio de Janeiro,                   
                  Rio de Janeiro, Brazil}                                     
\centerline{$^{4}$Institute of High Energy Physics, Beijing,                  
                  People's Republic of China}                                 
\centerline{$^{5}$Universidad de los Andes, Bogot\'{a}, Colombia}             
\centerline{$^{6}$Charles University, Center for Particle Physics,            
                  Prague, Czech Republic}                                     
\centerline{$^{7}$Institute of Physics, Academy of Sciences, Center           
                  for Particle Physics, Prague, Czech Republic}               
\centerline{$^{8}$Universidad San Francisco de Quito, Quito, Ecuador}         
\centerline{$^{9}$Institut des Sciences Nucl\'eaires, IN2P3-CNRS,             
                  Universite de Grenoble 1, Grenoble, France}                 
\centerline{$^{10}$CPPM, IN2P3-CNRS, Universit\'e de la M\'editerran\'ee,     
                  Marseille, France}                                          
\centerline{$^{11}$Laboratoire de l'Acc\'el\'erateur Lin\'eaire,              
                  IN2P3-CNRS, Orsay, France}                                  
\centerline{$^{12}$LPNHE, Universit\'es Paris VI and VII, IN2P3-CNRS,         
                  Paris, France}                                              
\centerline{$^{13}$DAPNIA/Service de Physique des Particules, CEA, Saclay,    
                  France}                                                     
\centerline{$^{14}$Universit{\"a}t Mainz, Institut f{\"u}r Physik,            
                  Mainz, Germany}                                             
\centerline{$^{15}$Panjab University, Chandigarh, India}                      
\centerline{$^{16}$Delhi University, Delhi, India}                            
\centerline{$^{17}$Tata Institute of Fundamental Research, Mumbai, India}     
\centerline{$^{18}$Seoul National University, Seoul, Korea}                   
\centerline{$^{19}$CINVESTAV, Mexico City, Mexico}                            
\centerline{$^{20}$FOM-Institute NIKHEF and University of                     
                  Amsterdam/NIKHEF, Amsterdam, The Netherlands}               
\centerline{$^{21}$University of Nijmegen/NIKHEF, Nijmegen, The               
                  Netherlands}                                                
\centerline{$^{22}$Institute of Nuclear Physics, Krak\'ow, Poland}            
\centerline{$^{23}$Joint Institute for Nuclear Research, Dubna, Russia}       
\centerline{$^{24}$Institute for Theoretical and Experimental Physics,        
                   Moscow, Russia}                                            
\centerline{$^{25}$Moscow State University, Moscow, Russia}                   
\centerline{$^{26}$Institute for High Energy Physics, Protvino, Russia}       
\centerline{$^{27}$Lancaster University, Lancaster, United Kingdom}           
\centerline{$^{28}$Imperial College, London, United Kingdom}                  
\centerline{$^{29}$University of Arizona, Tucson, Arizona 85721}              
\centerline{$^{30}$Lawrence Berkeley National Laboratory and University of    
                  California, Berkeley, California 94720}                     
\centerline{$^{31}$University of California, Davis, California 95616}         
\centerline{$^{32}$California State University, Fresno, California 93740}     
\centerline{$^{33}$University of California, Irvine, California 92697}        
\centerline{$^{34}$University of California, Riverside, California 92521}     
\centerline{$^{35}$Florida State University, Tallahassee, Florida 32306}      
\centerline{$^{36}$University of Hawaii, Honolulu, Hawaii 96822}              
\centerline{$^{37}$Fermi National Accelerator Laboratory, Batavia,            
                   Illinois 60510}                                            
\centerline{$^{38}$University of Illinois at Chicago, Chicago,                
                   Illinois 60607}                                            
\centerline{$^{39}$Northern Illinois University, DeKalb, Illinois 60115}      
\centerline{$^{40}$Northwestern University, Evanston, Illinois 60208}         
\centerline{$^{41}$Indiana University, Bloomington, Indiana 47405}            
\centerline{$^{42}$University of Notre Dame, Notre Dame, Indiana 46556}       
\centerline{$^{43}$Iowa State University, Ames, Iowa 50011}                   
\centerline{$^{44}$University of Kansas, Lawrence, Kansas 66045}              
\centerline{$^{45}$Kansas State University, Manhattan, Kansas 66506}          
\centerline{$^{46}$Louisiana Tech University, Ruston, Louisiana 71272}        
\centerline{$^{47}$University of Maryland, College Park, Maryland 20742}      
\centerline{$^{48}$Boston University, Boston, Massachusetts 02215}            
\centerline{$^{49}$Northeastern University, Boston, Massachusetts 02115}      
\centerline{$^{50}$University of Michigan, Ann Arbor, Michigan 48109}         
\centerline{$^{51}$Michigan State University, East Lansing, Michigan 48824}   
\centerline{$^{52}$University of Nebraska, Lincoln, Nebraska 68588}           
\centerline{$^{53}$Columbia University, New York, New York 10027}             
\centerline{$^{54}$University of Rochester, Rochester, New York 14627}        
\centerline{$^{55}$State University of New York, Stony Brook,                 
                   New York 11794}                                            
\centerline{$^{56}$Brookhaven National Laboratory, Upton, New York 11973}     
\centerline{$^{57}$Langston University, Langston, Oklahoma 73050}             
\centerline{$^{58}$University of Oklahoma, Norman, Oklahoma 73019}            
\centerline{$^{59}$Brown University, Providence, Rhode Island 02912}          
\centerline{$^{60}$University of Texas, Arlington, Texas 76019}               
\centerline{$^{61}$Texas A\&M University, College Station, Texas 77843}       
\centerline{$^{62}$Rice University, Houston, Texas 77005}                     
\centerline{$^{63}$University of Virginia, Charlottesville, Virginia 22901}   
\centerline{$^{64}$University of Washington, Seattle, Washington 98195}       

\end{center}

\normalsize

\vfill\eject

The observed symmetry between the lepton ($l$) and quark ($q$) sectors 
suggests the existence of a force connecting the two that is mediated by
particles that couple directly to both leptons and quarks, and are therefore 
known as leptoquarks ($LQ$).  Leptoquarks arise naturally as the vector bosons 
\cite{lq} or Higgs particles \cite{lqhiggs} of a Grand Unified Theory 
\cite{lq}; as composite particles \cite{lqcomp}; as techniparticles 
\cite{lqtechni}; or as R-parity violating supersymmetric particles 
\cite{rpvio}.
\par 
Leptoquarks carry both color and fractional electric charge.  The Fermilab 
Tevatron can produce pairs of leptoquarks through the strong process 
$p{\bar p} \rightarrow g \rightarrow LQ\overline{LQ} + X$ with a production 
cross section that is independent of the coupling for scalar leptoquarks, but 
not for vector leptoquarks.  In this study, we consider the specific cases 
of Yang-Mills coupling (YM), Minimal Coupling (MC), and the coupling resulting 
in the minimal cross section ($\sigma _{min}$) \cite{mc}.   
\par 
Decay between generations is theoretically possible; however, the limits from  
flavor-changing neutral currents imply that leptoquarks of low mass 
(${\cal O}$(Tev)) couple only within a single generation 
\cite{generation}.  Decays of leptoquark pairs result in one of three possible 
final states:  $l^{\pm}l^{\mp} qq$, $l^{\pm}\nu qq$, and $\nu\nu qq$.  This 
analysis \cite{thesis} studies the $\nu\nu qq$ final state, assuming 
BR($LQ \rightarrow \nu q$)=100\%.  In a previous analysis of this final state
\cite{d01a}, D\O\ set limits of $M_{LQ} >$ 79 GeV/c$^2$ for scalar 
leptoquarks, and $M_{LQ} >$ 145 GeV/c$^2$, 160 GeV/c$^2$, and 205 GeV/c$^2$, 
for vector leptoquarks with couplings that yield the minimum cross section 
($\sigma _{min}$), Minimal Couplings (MC), and Yang-Mills (YM) couplings, 
respectively \cite{d01a}.  The analysis presented here uses 10 times more data 
than the previous analysis.  The CDF collaboration has conducted a search for 
second and third generation leptoquarks with BR($LQ \rightarrow \nu q$)=100\% 
and set mass limits of 123 (148) GeV/c$^2$ for second (third) generation 
scalar leptoquarks and 171 (199) GeV/c$^2$ and 222 (250) GeV/c$^2$ for second 
(third) generation vector leptoquarks with MC and YM couplings, respectively 
\cite{cdfg23}.   
\par 
The D\O\ detector \cite{detector} consists of three major subsystems:  An  
inner detector for tracking charged particles; a uranium-liquid argon  
calorimeter for measuring electromagnetic and hadronic showers, and a muon  
spectrometer.  The jets measured with the calorimeter have an energy  
resolution of approximately $\sigma(E)$=0.8$\sqrt{E}$ ($E$ in GeV).  We  
measure the missing transverse energy (\met) by summing the calorimeter energy 
in the direction transverse to the beam.  The measurement has a resolution of 
$\sigma$=1.08 GeV + 0.019($\Sigma |E_T|$) ($E_T$ in GeV). 
\par 
We use an event sample defined by the selection criteria:  2 jets with $E_T >$ 
50 GeV; \met $>$ 40 GeV; $\Delta\phi (jet,~$\met$) >$ 30$^{\circ}$; and 
$\Delta {\cal R} (jet,~jet) >$ 1.5, where ${\Delta \cal R}=
\sqrt{(\Delta\eta)^2 + (\Delta\phi)^2}$, $\eta$ is the jet pseudo-rapidity, 
and $\phi$ is the jet azimuthal angle.  These criteria select events with high 
trigger efficiency.  Backgrounds arising from $W$ or $Z$ boson production are 
reduced by rejecting events with isolated muons or highly electromagnetic 
jets.  We reduce cosmic ray backgrounds by rejecting events with jets 
containing little electromagnetic activity.  The integrated luminosity after 
removing events corrupted by accelerator noise and detector malfunctions is 
85.2 $\pm$ 3.7 pb$^{-1}$.
\par 
The backgrounds in the sample consist of events with jets produced in 
association with a $W$ or a $Z$ boson, and events from top quark and multijet 
production.  We use Monte Carlo generators to simulate the topologies of 
events with $W$ or $Z$ bosons or top quarks, and a GEANT simulation of the 
detector to predict the acceptance of these events.
\par
The $W$ and $Z$ backgrounds consist of processes containing only neutrinos and 
jets ($W \rightarrow \tau _h\nu +jet,~Z \rightarrow \nu\nu ~+~2~jets$), 
processes with unobserved charged leptons ($W \rightarrow l^{\pm} \nu 
~+~2~jets,~Z \rightarrow \mu\mu ~+~2~jets, ~Z \rightarrow \tau _h \tau _l 
~+~jet$), and processes in which an electron is misidentified as a jet 
($W \rightarrow e\nu ~+~jet, ~W \rightarrow \tau _e \nu ~+~jet$).  We use the 
\small PYTHIA \normalsize Monte Carlo generator \cite{pythia} to predict the 
acceptances of the $W/Z$ + jet processes, and the \small VECBOS \normalsize 
Monte Carlo generator \cite{vecbos} to predict the acceptances of the $W/Z$ + 
2 jets processes.  We scale the generator cross sections to match the cross 
sections measured using the $W$ and $Z$ electronic decays.   
\par 
The top background consists of $t{\bar t}$, $t{\bar b}$, and ${\bar t}b$  
production, where the top quark decays to an unobserved charged lepton, a 
neutrino, and a jet.  We use the D\O\ measured cross section for $t{\bar t}$ 
production \cite{ttbar} and the calculated next-to-leading order cross section 
for the other processes \cite{singlet}.  We use the \small HERWIG \normalsize 
generator \cite{herwig} to predict the acceptance of the $t{\bar t}$ process, 
and \small CompHEP \normalsize \cite{comphep} to predict the acceptances of 
the $t{\bar b}$ and ${\bar t}b$ processes. 
\par 
The multijet background arises primarily from 2 sources:  Vertex 
mismeasurement and jet energy loss.  To reduce the number of events with 
mismeasured vertices, we use the central drift chamber (CDC) to associate 
charged tracks with each central high $p_T$ jet.  The tracks are used to 
determine the jet vertex position which is required to be no further than 15 
cm from the event vertex position.  The latter is determined using all of the 
tracks in the event.  We reduce the number of events with significant jet 
energy loss by requiring that the angle between the \met\ and the jet with the 
second highest measured $p_T$ be greater than 60$^{\circ}$. 
\par 
To predict the multijet background remaining in our sample, we use the sample  
of events whose jet vertex position deviates by 15 cm to 50 cm from the event 
vertex position.  We normalize this sample to the event sample using a 
multijet dominated sample ($\Delta\phi (jet~2,~$\met$) < 60 ^{\circ}$).  The 
upper bound of 50 cm provides the best agreement between the background 
prediction and the data for events with \met\ between 30 GeV and 40 GeV, which 
is dominated by multijet events (Table \ref{tbl:qcd3040}).  Changing this 
value to 100 cm increases the multijet background prediction by 22\% in this 
region, which we take as an estimate of the systematic error of the 
method.  Table \ref{tbl:datbd} shows the total expected background and the 
observed number of events for the 2 jets + \met\ data sample. 
\par 
To model the characteristics of leptoquark production, we use scalar  
leptoquark events generated with \small PYTHIA \normalsize and vector  
leptoquark events generated with \small CompHEP\normalsize .  The cross  
sections for scalar leptoquark production have been calculated to
next-to-leading order \cite{kraemer}, while those for vector leptoquark  
production have been calculated to leading order \cite{vlqxsec}.  The  
calculations use a renormalization and factorization scale of $\mu$=M$_{LQ}$, 
with theoretical uncertainties estimated by changing the scale to 
$\mu$=M$_{LQ}$/2 and $\mu$=2M$_{LQ}$.  We use the lower cross section 
($\mu$=2M$_{LQ}$) in our optimization and in determining our mass limits.   
\par
\begin{figure}[!htbp]  
\begin{minipage}[htb]{8.0cm} 
\epsfysize = 8.0cm  
\epsffile{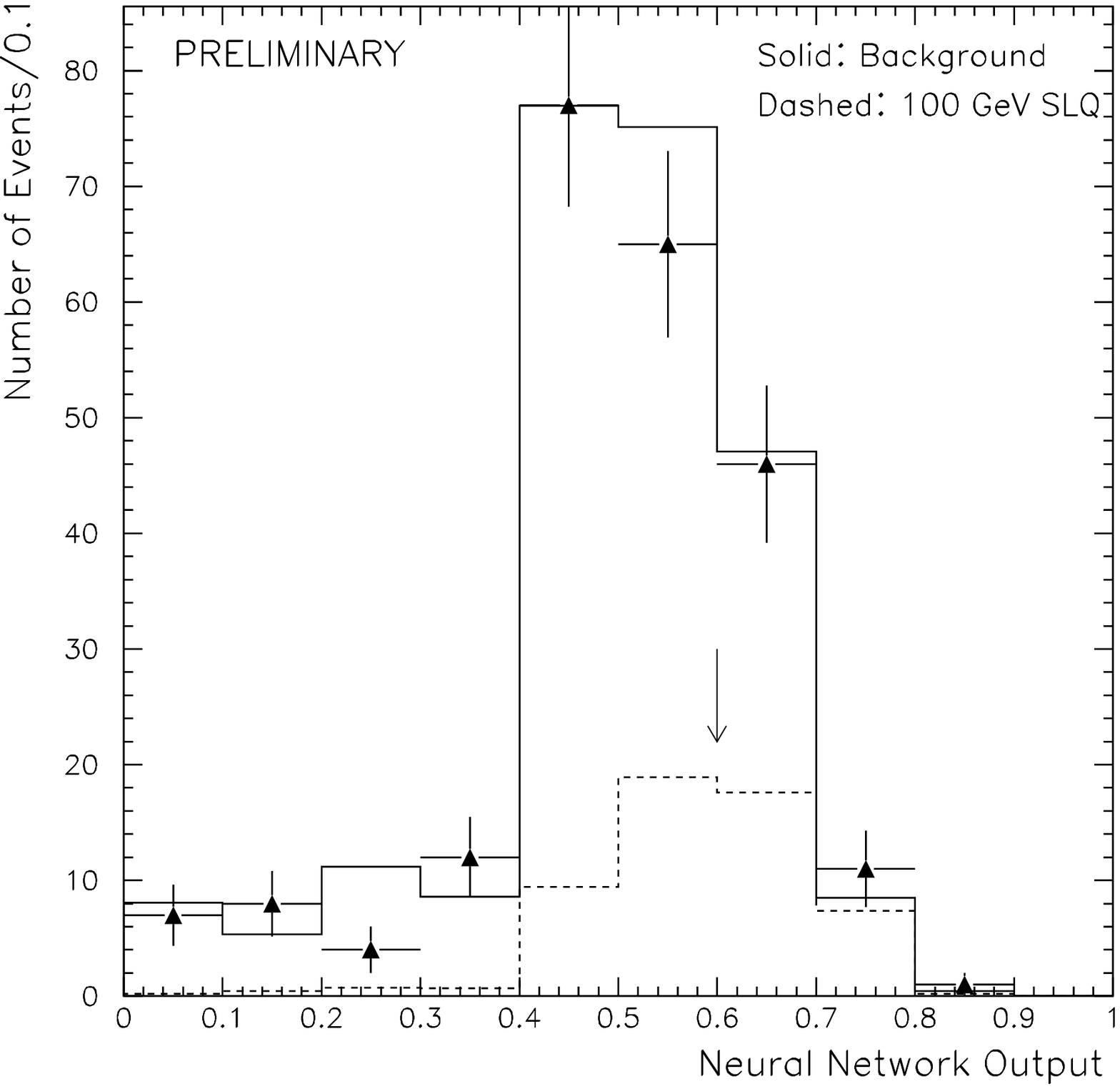} 
\end{minipage} 
\begin{minipage}[htb]{8.0cm} 
\epsfysize = 8.0cm  
\epsffile{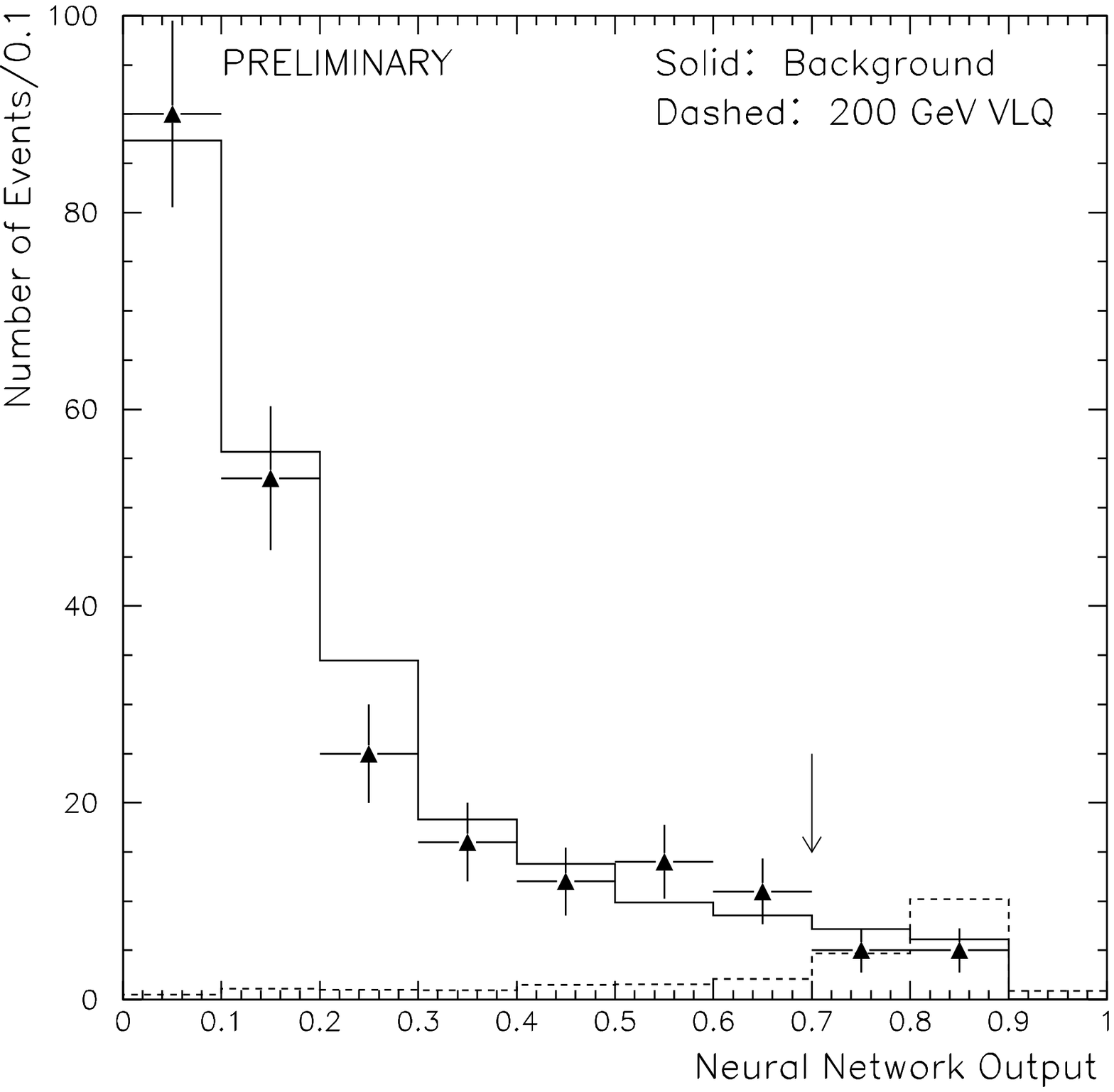} 
\end{minipage} 
\caption{The neural network output for the data, for background (solid), and  
for leptoquarks (dashed).  We show the optimization for 100 GeV/c$^2$ scalar  
leptoquarks (left) and 200 GeV/c$^2$ vector leptoquarks with Minimal Coupling
(right).  We remove events to the left of the arrows.} 
\label{fig:nnout} 
\end{figure}

 
The analysis is optimized for the production of 100 GeV/c$^2$ scalar 
leptoquarks and 200 GeV/c$^2$ vector leptoquarks, since leptoquarks with 
either of these masses would give a $\sim$2$\sigma$ excess if they exist.  We 
use the \small JETNET \normalsize \cite{jetnet} neural network program, with 
the \met\ and $\Delta\phi (jet,~jet)$ distributions as inputs for scalar 
leptoquarks, and the \met\ and second jet $p_T$ distributions as inputs for 
vector leptoquarks.  We show the neural network outputs and the chosen cuts 
for both of these masses in Fig. \ref{fig:nnout}.  The cuts are chosen to 
maximize the inverse of the fractional error in the signal:
 
\begin{center}
$n_{\sigma}=\frac{N_{lq}}{\sqrt{N_{lq}+N_{background}+\Delta N_{lq}^2+\Delta N_{background}^2}}$,
\end{center}  
 
\noindent 
where $N_{lq}$ and $N_{background}$ are the number of signal and background
events, respectively, and $\Delta N_{lq}$ and $\Delta N_{background}$ are the
associated uncertainties.  We show the numbers of events after these cuts in 
Table \ref{tbl:nn}. 

\begin{table}[hptb] 
\begin{center} 
\begin{tabular}{|cc|} 
Event Sample & Number of Events \\ 
\hline 
\hline Multijet & 162.8 $\pm$ 23.7\\ 
\hline W, Z, and top & 51.9 $\pm$ 7.0 \\ 
\hline Total background & 214.7 $\pm$ 24.7 \\ 
\hline 
\hline Data & 224 \\ 
\end{tabular} 
\end{center} 
\caption{The expected and observed numbers of events in the multijet  
dominated sample of \met\ between 30 GeV and 40 GeV.} 
\label{tbl:qcd3040} 
\end{table} 

\begin{table}[hptb] 
\begin{center} 
\begin{tabular}{|cc|} 
Background & Number of Events \\ 
\hline 
\hline Multijet & 58.8 $\pm$ 14.1 $\pm$ 12.9\\ 
\hline ($W \rightarrow e\nu )~+~jet$ & 51.9 $\pm$ 7.0 $^{+13.7} \! \! \! \! \! \! \! \! \! \! \! \! \! _{-8.9}$ \\ 
\hline ($W \rightarrow \tau\nu )~+~jet$ & 46.3 $\pm$ 5.0 $^{+8.9} \! \! \! \! \! \! \! \! \! \! _{-7.7}$ \\ 
\hline ($Z \rightarrow \nu\nu )~+~2~jets$ & 36.1 $\pm$ 7.7 $^{+9.0} \! \! \! \! \! \! \! \! \! \! \! _{-5.5}$ \\ 
\hline ($W \rightarrow \mu\nu )~+~2~jets$ & 18.7 $\pm$ 3.5 $^{+4.2} \! \! \! \! \! \! \! \! \! \! _{-3.7}$ \\ 
\hline $t{\bar t} \rightarrow$ $l^{\pm}\nu ~+~4~jets$ & 10.6 $\pm$ 2.0 $\pm$ 2.3 \\ 
\hline ($W \rightarrow e\nu )~+~2~jets$ & 8.3 $\pm$ 2.5 $^{+2.0} \! \! \! \! \! \! \! \! \! \! _{-2.5}$ \\ 
\hline ($W \rightarrow \tau\nu )~+~2~jets$ & 5.6 $\pm$ 1.7 $^{+1.4} \! \! \! \! \! \! \! \! \! \! _{-0.8}$ \\ 
\hline $tb \rightarrow$ $l^{\pm}\nu ~+~2~jets$ & 2.0 $\pm$ 0.3 $\pm$ 0.2 \\  
\hline ($Z \rightarrow \tau\tau )~+~jet$ & 2.0 $\pm$ 0.4 $^{+0.6} \! \! \! \! \! \! \! \! \! \! _{-0.3}$ \\ 
\hline ($Z \rightarrow \mu\mu )~+~2~jets$ & 1.7 $\pm$ 0.4 $^{+0.4} \! \! \! \! \! \! \! \! \! \! _{-0.3}$ \\ 
\hline Total background & 242.0 $\pm$ 18.9 $^{+23.3} \! \! \! \! \! \! \! \! \! \! \! \! \! _{-19.0}$ \\ 
\hline 
\hline Data & 231 \\ 
\end{tabular} 
\end{center} 
\caption{The expected and observed numbers of events in the 2 jets + \met\ 
sample.} 
\label{tbl:datbd} 
\end{table} 

\begin{table}[hptb] 
\begin{center} 
\begin{tabular}{|c|c|c|c|c|} 
Leptoquark & $N_{data}$       & $N_{background}$      & $n_{\sigma}$ & $\sigma ^{95\%}$ (pb)\\ 
\hline 100 GeV/c$^2$ Scalar & 58 & 56.0 $^{+8.1} \! \! \! \! \! \! \! \! \! \! \! _{-8.2}$ & +2.1 & 10.8 \\ 
\hline 200 GeV/c$^2$ Vector (MC) & 10 & 13.3 $^{+2.8} \! \! \! \! \! \! \! \! \!  _{-2.6}$ & +2.6 & 0.60 \\ 
\end{tabular} 
\end{center} 
\caption{The data, the expected background, the number of $\sigma$ excess that 
would be observed in the presence of a signal, and the 95\% confidence level  
cross section limit.} 
\label{tbl:nn} 
\end{table} 

\begin{figure}[htb] 
\begin{minipage}[htb]{8.0cm} 
\epsfysize = 8.0cm  
\epsffile{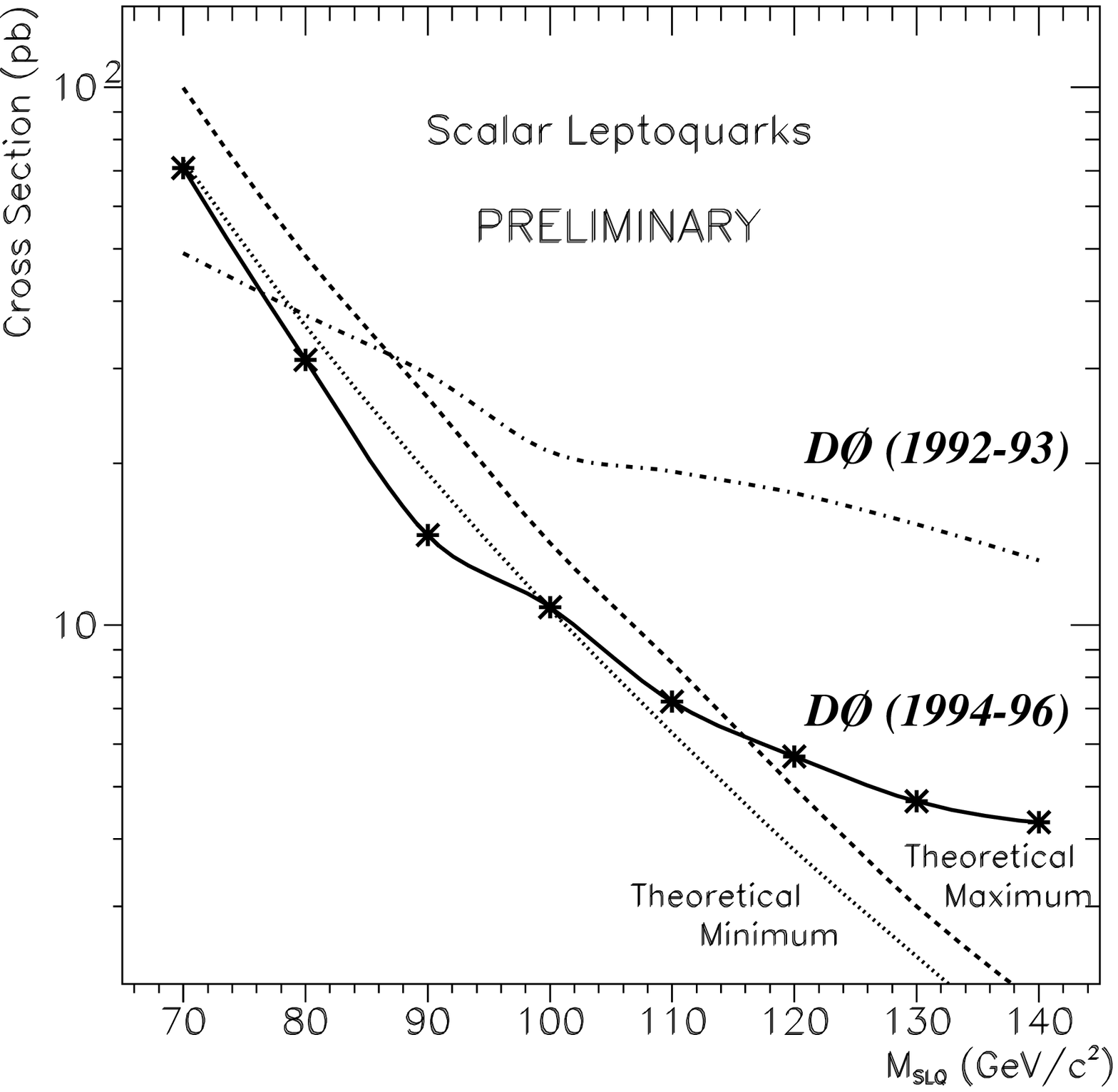} 
\end{minipage} 
\begin{minipage}[htb]{8.0cm} 
\epsfysize = 8.0cm  
\epsffile{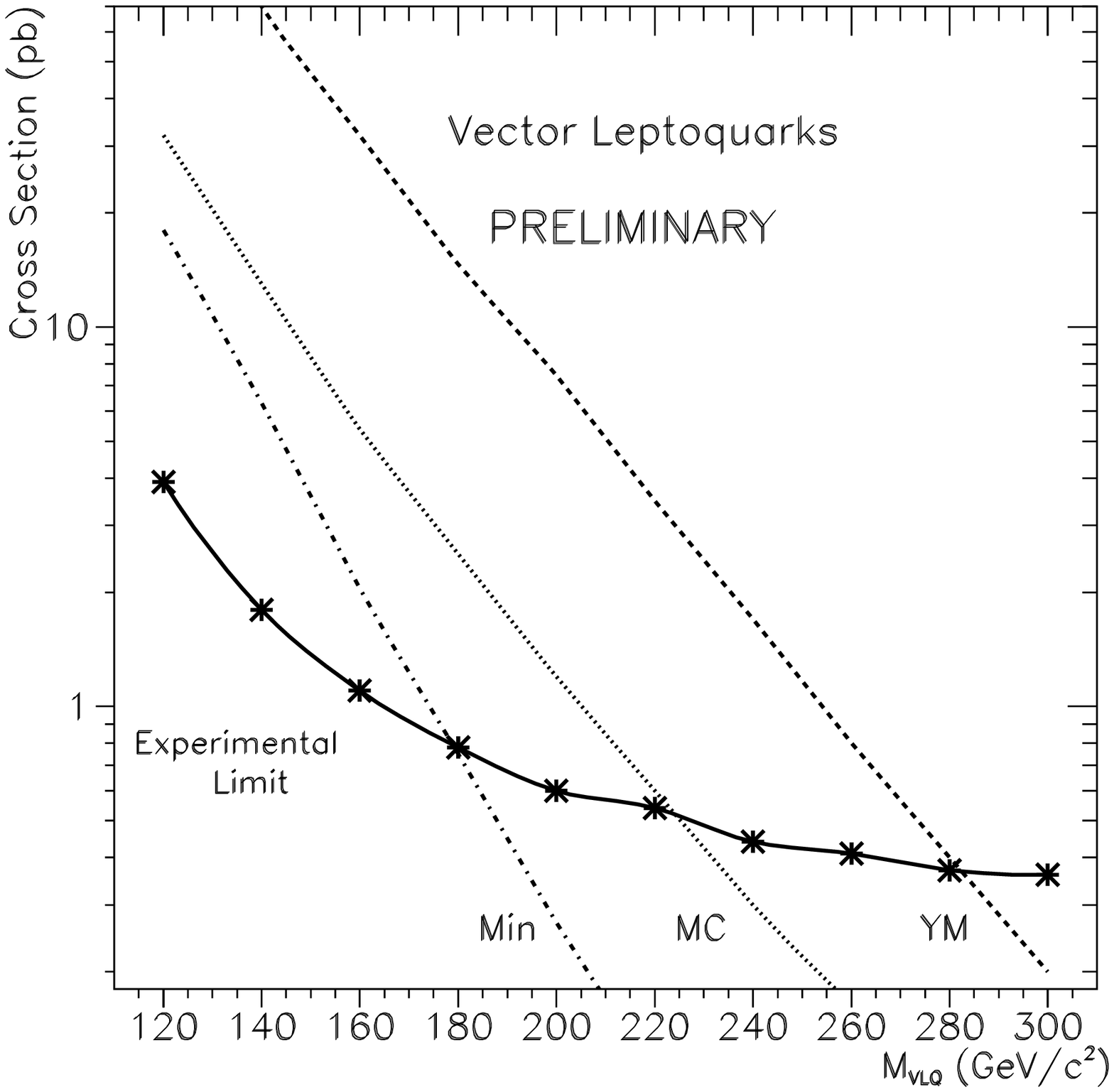} 
\end{minipage} 
\caption{The 95\% confidence level cross section limits as a function of  
leptoquark mass.  We show the mass limits for scalar (left) and vector (right) 
leptoquarks.} 
\label{fig:lqlim} 
\end{figure}

\begin{figure}[hptb] 
\epsfxsize = 9.5cm 
\centerline{\epsffile{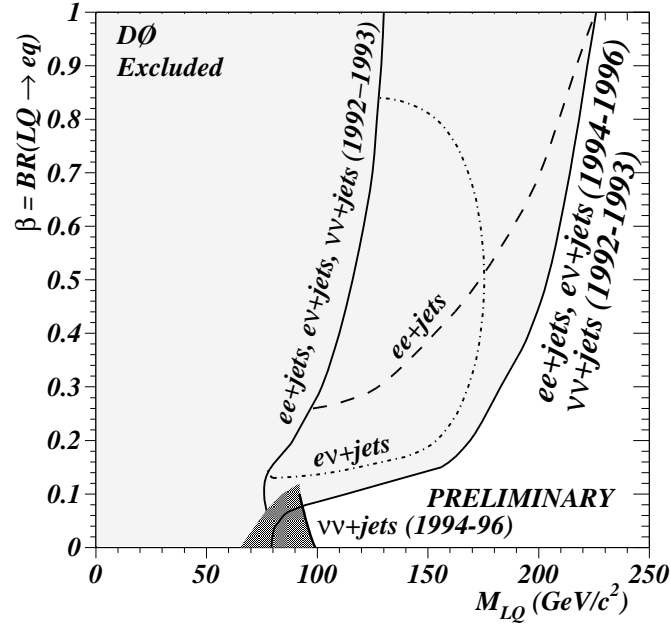}} 
\vskip -0.1in 
\caption{The D\O\ excluded mass vs. BR($LQ \rightarrow eq$) region for first  
generation scalar leptoquarks.  The dark region is excluded by this analysis.} 
\label{fig:slqbr} 
\end{figure}

\begin{figure}[hptb] 
\begin{minipage}[hptb]{8.0cm} 
\epsfysize = 8.0cm  
\epsffile{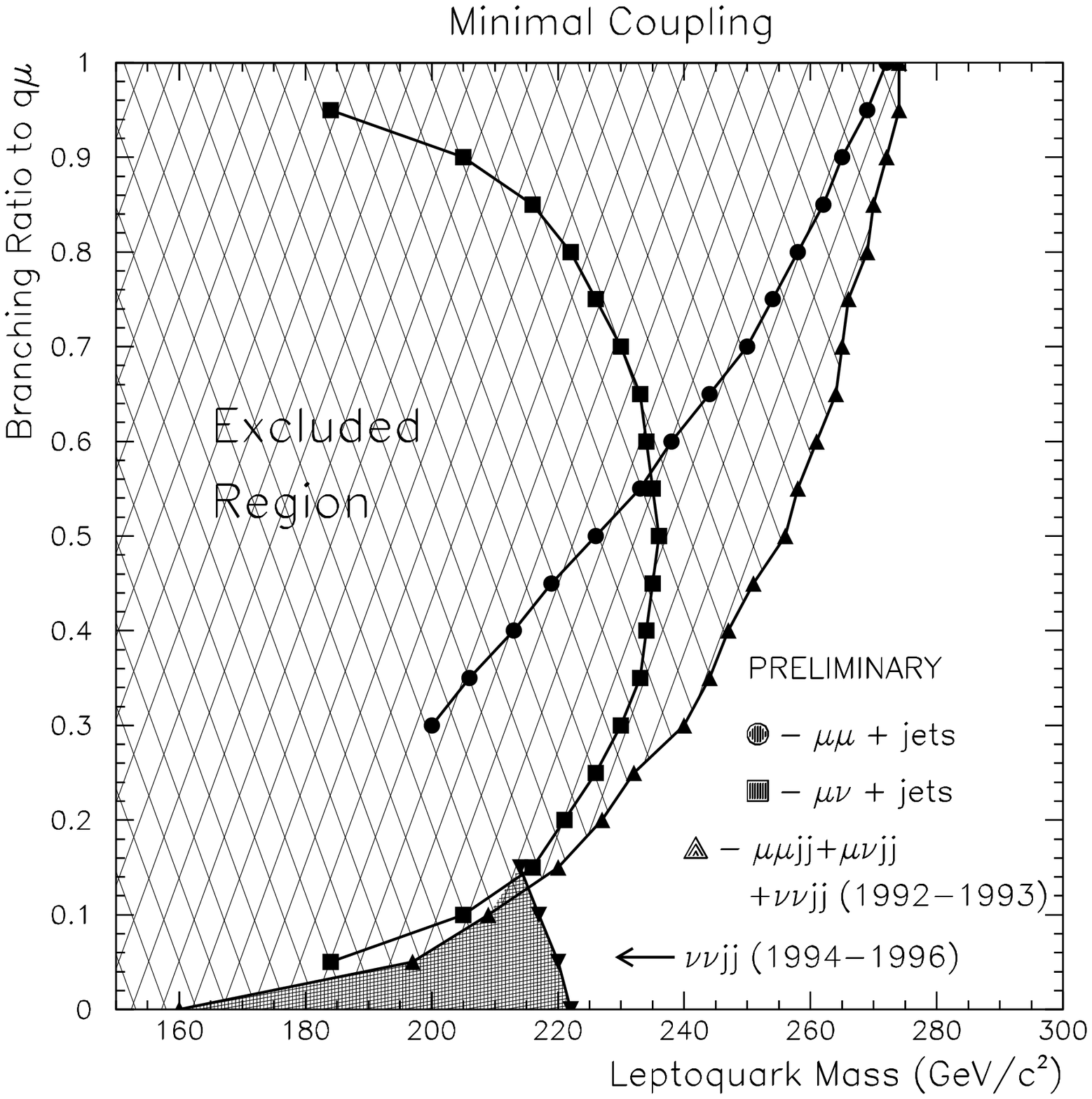} 
\end{minipage} 
\begin{minipage}[hptb]{8.0cm} 
\epsfysize = 8.0cm     
\epsffile{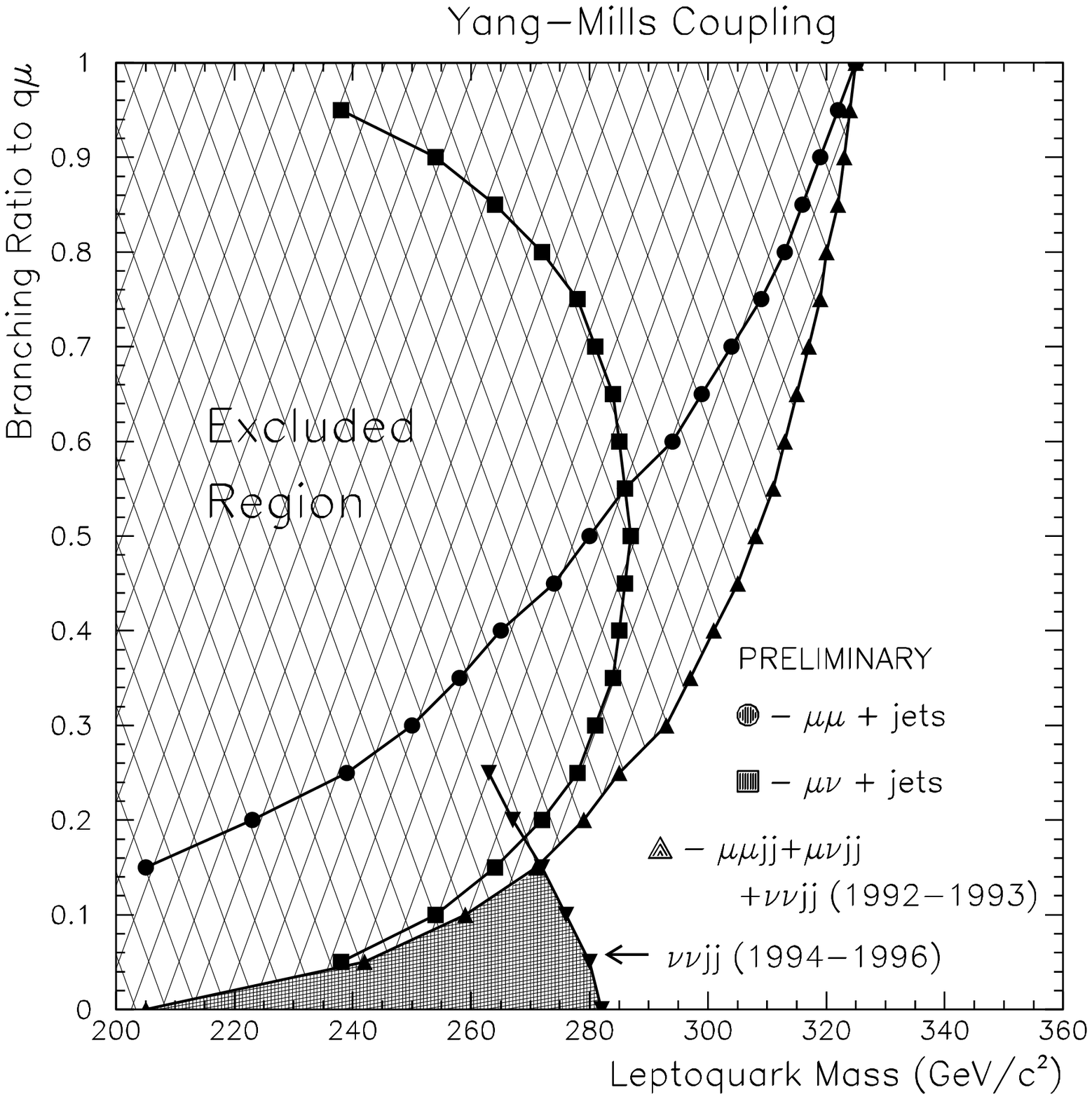} 
\end{minipage} 
\caption{The D\O\ excluded mass vs. BR($LQ \rightarrow \mu q$) region for second  
generation MC (left) and YM (right) vector leptoquarks.  The dark regions are
excluded by this analysis.} 
\label{fig:vlqbr} 
\end{figure}

After applying the optimal cuts, we find that the observed number of events 
is consistent with the expected background, and that, consequently, we have
found no evidence for leptoquark production.  This null result yields the 
95\% confidence level cross section limit, as a function of leptoquark mass, 
shown in Fig. \ref{fig:lqlim}.  We calculate the limit using a Bayesian method 
with a flat prior for the signal and Gaussian priors for background and 
acceptance uncertainties.  The corresponding mass limits are 99 GeV/c$^2$ for 
scalar leptoquarks, and 178 GeV/c$^2$, 222 GeV/c$^2$, and 282 GeV/c$^2$ for 
vector leptoquarks with couplings corresponding to the minimum cross section 
$\sigma _{min}$, Minimal Coupling, and Yang-Mills coupling, respectively.  We 
summarize the various D\O\ mass limits as a function of 
BR($LQ \rightarrow l^{\pm}q$) for first generation scalar leptoquarks in Fig. 
\ref{fig:slqbr} \cite{d01a} and for second generation MC and YM vector 
leptoquarks \cite{d0g2} in Fig. \ref{fig:vlqbr}.  We note that the gap at low 
values of BR($LQ \rightarrow l^{\pm}q$) has been closed by this analysis.  

\section*{Acknowledgements}
\label{sec:ack}
%
%
We thank the staffs at Fermilab and collaborating institutions, and
acknowledge support from the Department of Energy and National Science
Foundation (USA), Commissariat \` a L'Energie Atomique and
CNRS/Institut National de Physique Nucl\'eaire et de Physique des
Particules (France), Ministry for Science and Technology and Ministry
for Atomic Energy (Russia), CAPES and CNPq (Brazil), Departments of
Atomic Energy and Science and Education (India), Colciencias
(Colombia), CONACyT (Mexico), Ministry of Education and KOSEF (Korea),
CONICET and UBACyT (Argentina), The Foundation for Fundamental
Research on Matter (The Netherlands), PPARC (United Kingdom), Ministry
of Education (Czech Republic), and the A.P.~Sloan Foundation.

\end{document}